\documentclass[twocolumn]{aastex631}
\shorttitle{$\rm CO(2\to1)$ detection in a $z=6$ galaxy}
\shortauthors{Zavala et al.}
\graphicspath{{./}{figures/}}

\begin{document}

\title{Probing cold gas in a massive, compact star-forming galaxy at $\mathbf{z=6}$}

\author[0000-0002-0786-7307]{Jorge A. Zavala}
\affiliation{National Astronomical Observatory of Japan, 2-21-1 Osawa, Mitaka, Tokyo 181-8588, Japan}
\affiliation{Department of Astronomy, The University of Texas at Austin, 2515 Speedway Blvd Stop C1400, Austin, TX 78712, USA}
\author[0000-0002-0930-6466]{Caitlin M. Casey}
\affiliation{Department of Astronomy, The University of Texas at Austin, 2515 Speedway Blvd Stop C1400, Austin, TX 78712, USA}
\author[0000-0003-3256-5615]{Justin Spilker}
\affiliation{Department of Physics and Astronomy and George P. and Cynthia Woods Mitchell Institute for Fundamental Physics and Astronomy, Texas A\&M University, 4242 TAMU, College Station, TX 77843-4242}
\author[0000-0001-9728-8909]{Ken-ichi Tadaki}
\affiliation{National Astronomical Observatory of Japan, 2-21-1 Osawa, Mitaka, Tokyo 181-8588, Japan}
\author[0000-0002-0498-5041]{Akiyoshi Tsujita}
\affiliation{Institute of Astronomy, Graduate School of Science, The University of Tokyo, 2-21-1 Osawa, Mitaka, Tokyo 181-0015, Japan}
\author[0000-0002-6184-9097]{Jaclyn Champagne}
\affiliation{Department of Astronomy, The University of Texas at Austin, 2515 Speedway Blvd Stop C1400, Austin, TX 78712, USA}
\author[0000-0002-2364-0823]{Daisuke Iono}
\affiliation{National Astronomical Observatory of Japan, 2-21-1 Osawa, Mitaka, Tokyo 181-8588, Japan}
\author[0000-0002-4052-2394]{Kotaro Kohno}
\affiliation{Institute of Astronomy, Graduate School of Science, The University of Tokyo, 2-21-1 Osawa, Mitaka, Tokyo 181-0015, Japan}
\author[0000-0003-0415-0121]{Sinclaire Manning}
\affiliation{Department of Astronomy, University of Massachusetts, Amherst, MA 01003 USA}
\author[0000-0003-4229-381X]{Alfredo Monta\~na}
\affiliation{Instituto Nacional de Astrof\'isica, \'Optica y Electr\'onica, Luis Enrique Erro 1, CP 72840, Tonantzintla, Puebla, Mexico}



\begin{abstract}
Observations of low order $^{12}\text C^{16}\text O$ transitions represent the most direct way to study galaxies' cold molecular gas, the fuel of star formation. Here we present the first detection of CO($J$=2$\to$1) in a galaxy lying on the main-sequence of star-forming galaxies at $z>6$. 
Our target, G09-83808 at $z=6.03$, has a short depletion time-scale of $\tau_{\rm dep}\approx50\rm\,Myr$ and a relatively low gas fraction of $\rm M_{\rm gas}/M_\star\approx0.30$ that contrasts with those measured for lower redshift main-sequence galaxies. We conclude that this galaxy is undergoing a starburst episode with a high star formation efficiency that might be the result of gas compression within its compact rotating disk.  Its starburst-like
nature is further supported by its high star formation rate surface density, thus favoring the use of
the Kennicutt-Schmidt relation as a more precise diagnostic diagram. Without further significant gas accretion, this galaxy would become a compact, massive quiescent galaxy at $z\sim5.5$.

In addition, we find that the calibration for estimating ISM masses from dust continuum emission satisfactorily reproduces the gas mass derived from the CO(2$\to$1) transition (within a factor of $\sim2$). This is in line with previous studies claiming a small redshift evolution in the gas-to-dust ratio of massive, metal-rich galaxies.


In the absence of gravitational amplification, this detection would have required of order $1000\,$h of observing time. The detection of cold molecular gas in unlensed star-forming galaxies at high redshifts is thus prohibitive with current facilities and requires a ten-fold improvement in sensitivity, such as that envisaged for the ngVLA. 

\end{abstract}



\section{Introduction} \label{sec:intro}


The evolution of the star formation activity and, consequently, the cosmic mass assembly history and the metal enrichment of the universe are tightly linked to the molecular gas content of galaxies throughout cosmic time (e.g. \citealt{Walter2020a}). {\it How much gas was present in these `cosmic ecosystems' at different epochs} is a fundamental question in galaxy formation and evolution studies. 

$\text H_2$, the most abundant molecule in the Universe, lacks dipole moments and therefore is a very inefficient radiator; thus the cool molecular gas' best observable tracer is thought to be carbon monoxide ($^{12}\text C^{16}\text O$; hereafter, CO), the second most abundant molecule in the universe.
Since a few years after its first astronomical detection ---
half a century ago (\citealt{Wilson1970a,Penzias1971a}),  it was clear that this rotational emission line was a useful tool to constrain the molecular gas content in extragalactic sources and to map its distribution and kinematics (e.g. \citealt{Rickard1975a,Solomon1975a}) since it is easily excited even at very low temperatures. 

The estimation of the total molecular gas from the CO line luminosity relies on the CO–to–H$_2$ conversion factor, $\alpha_{\text{CO}}$, usually calibrated for the $\rm CO(1\to0)$  transition\footnote{This relation has the form of $\text M_{\text mol}=\alpha_{\text CO}L'_{\text CO}$; where $\text M_{\text mol}$ has units of $\text M_\odot$, and $L'_{\text CO}$, the CO line luminosity, is expressed in units of $\text K\,\text{km}\,\text s^{-1}\,\rm pc^2$.} (see review by \citealt{Bolatto2013a}). When only higher order transitions are available, an assumption on the excitation ladder (or a modeling if several lines are available) has to be done as an additional step to first infer the CO ground transition line luminosity. 
The shape of the spectral line energy distribution (SLED), however, depends on the excitation of the CO molecules (and therefore on the temperature and density of the gas), which might vary significantly between galaxies and even between regions of the same galaxy. This introduces significant uncertainty when inferring the low-J line luminosities from  high-J transitions, at the level of making gas mass estimations unreliable, particularly when only one 
high-order line is available\footnote{The CO(6$\to$5)/CO(1$\to$0) flux line ratios found in submillimeter-selected galaxies vary, for example, by up to factors of six, or even up to a factor of ten if quasars are included (see review by \citealt{Carilli2013a}).}. This is particularly true at $z>1$ given the sensitivity and frequency coverage of current facilities and the typical SLED of star-forming galaxies, which bias the observations towards the brighter high-J CO emission lines. 

Despite these difficulties, hundreds of CO studies in galaxies from $z=0$ to $z\approx3-4$ have been published to-date, allowing us to constrain the physical properties and redshift evolution of galaxies' interstellar medium (see recent review by \citealt{Tacconi2020a} and references therein). Moreover, these observations (in addition to other dynamical studies) have been used to calibrate alternative tracers of the molecular gas component such as  far–infrared (FIR) atomic fine structure lines (e.g. \citealt{Valentino2018a,Dessauges-Zavadsky2020a}) and dust continuum emission (e.g. \citealt{Magdis2012a,Scoville2016a}).

At higher redshifts, however,
our understanding of the physical conditions of the cold star-forming interstellar medium is much more limited. And, while recent observational programs, such as the ALPINE survey (\citealt{Bethermin2020a,LeFevre2020}) have made a significant leap forward in our understanding of these quantities up to $z\sim5.5$ (e.g. \citealt{Dessauges-Zavadsky2020a}), primarily through observations of [CII], the number of galaxies with low-J CO detections at these high redshifts is limited to a handful (e.g. \citealt{Strandet2017a,Pavesi2018}). At $z>6$, the situation is even more critical. The only
detections of $\rm CO(1\to0)$ and $\rm CO(2\to1)$  come from a very extreme starburst galaxy, whose star formation rate is thought to be among the highest observed at any epoch (\citealt{Riechers2013a}). The gas properties of normal star-forming galaxies at $z>6$ are thus completely unexplored.
 
The reason behind this relies mainly on the limitations of existing instrumentation since the current facilities covering the radio wavebands at which  these  low-J CO transitions are redshifted ($\nu_{\rm obs}\approx15-35\,\rm GHz$) lack the required sensitivity to detect the  expected CO emission from typical star-forming galaxies at $z\gtrsim5$. Future large radio arrays are currently being designed to specifically address this issue. The Next-Generation Very Large Array (ngVLA), for example, will provide an order-of-magnitude improvement in depth and area (\citealt{Carilli2015a,Murphy2018a,Selina2018a}) allowing us to conduct surveys of cold gas from these critical low-J transitions in normal galaxies within the Epoch of Reionzation (\citealt{Casey2015a,Casey2018c,Decarli2018a}). 

Nevertheless, such facilities are not expected to be operational until the end of this decade, at the earliest. Furthermore, the capability of such a facility to detect cold gas at high redshifts ($z>5$) has been called into question due to the potentially strong dimming effects caused by the Cosmic Microwave Background (e.g. \citealt{daCunha2013a}). 

Here, we exploit the gravitational amplification effect, to demonstrate the feasibility of detecting cold gas in normal galaxies within the Epoch of Reionization via the first detection of $\rm CO(2\to1)$  in a main-sequence star-forming galaxy at $z=6$. The detection of this line, from which the ground $\rm CO(1\to0)$  line luminosity can be estimated with little extrapolation,
enable us to characterize the cold ISM in a galaxy formed $<1\,$Gyr after the Big Bang. This allows us to test the applicability of scaling relations derived for lower redshift galaxies and test whether or not other widely used methods to infer gas masses are also valid for this $z=6$ system.

This work provides us the first insights of what is expected to come with future facilities such as the ngVLA (\citealt{Carilli2015a}) and the new Band-1 ALMA detectors (\citealt{Huang2016a}).

In this manuscript, we assume $H_0=67.3\,\rm km\,s^{-1}\,Mpc^{-1}$ and $\Omega_\lambda=0.68$ (\citealt{Planck2016a}).

\section{Target and observations}

\subsection{G09-83808: a massive, main-sequence star forming galaxy at $z=6$?}\label{sec:target}
The target, G09-83808, was first detected by the {\it Herschel} space telescope and identified as a high-redshift galaxy candidate in \citet{Ivison2016a} due to its red far-infrared colors (i.e. $S_{\rm 250\mu m}<S_{\rm 350\mu m}<S_{\rm 500\mu m}$).
The source was then followed-up with several telescopes including the Atacama Large submillimeter/Millimeter Array (ALMA), the James Clerk Maxwell Telescope (JCMT), the Large Millimeter Telescope (LMT), NOEMA, {\it Spitzer}, and the Submillimeter Array (SMA). The galaxy's redshift was determined to be $z=6.0269\pm0.0006$ through the detection of multiple emission lines (including $\rm CO(5\to4)$ , $\rm CO(6\to5)$ , and $\rm H_2O(2_{11}-2_{02})$) in its LMT 3\,mm spectrum and a subsequent detection of the [CII]($\rm^2P_{3/2}-^2P_{1/2}$) transition with the SMA (\citealt{Zavala2018a}; see also \citealt{Fudamoto2017a}).
\citet{Zavala2018a} also presents high-angular resolution ALMA observations of the dust continuum emission at $\lambda_{\rm obs}\sim890\,\mu\rm m$. The high resolution dust continuum observations were used for modelling the gravitational lensing effect
in the {\it uv} plane 
using the {\sc visilens} code (\citealt{Spilker2016a}). The best-fit gravitational  magnification of the source was found to be $\mu_{890\mu\rm m}=9.3\pm1.0$. 

Using this magnification factor and the combined {\it Herschel}/SPIRE (250, 350, and $500\,\mu\rm m$), JCMT/SCUBA-2 $850\,\mu\rm m$, and LMT/AzTEC 1.1\,mm photometry, an intrinsic star formation rate (SFR) of  $380\pm50\,\rm M_\odot\,\rm yr^{-1}$ was derived in \citet{Zavala2018a}.

\begin{figure}
\centering
\includegraphics[width=0.5\textwidth]{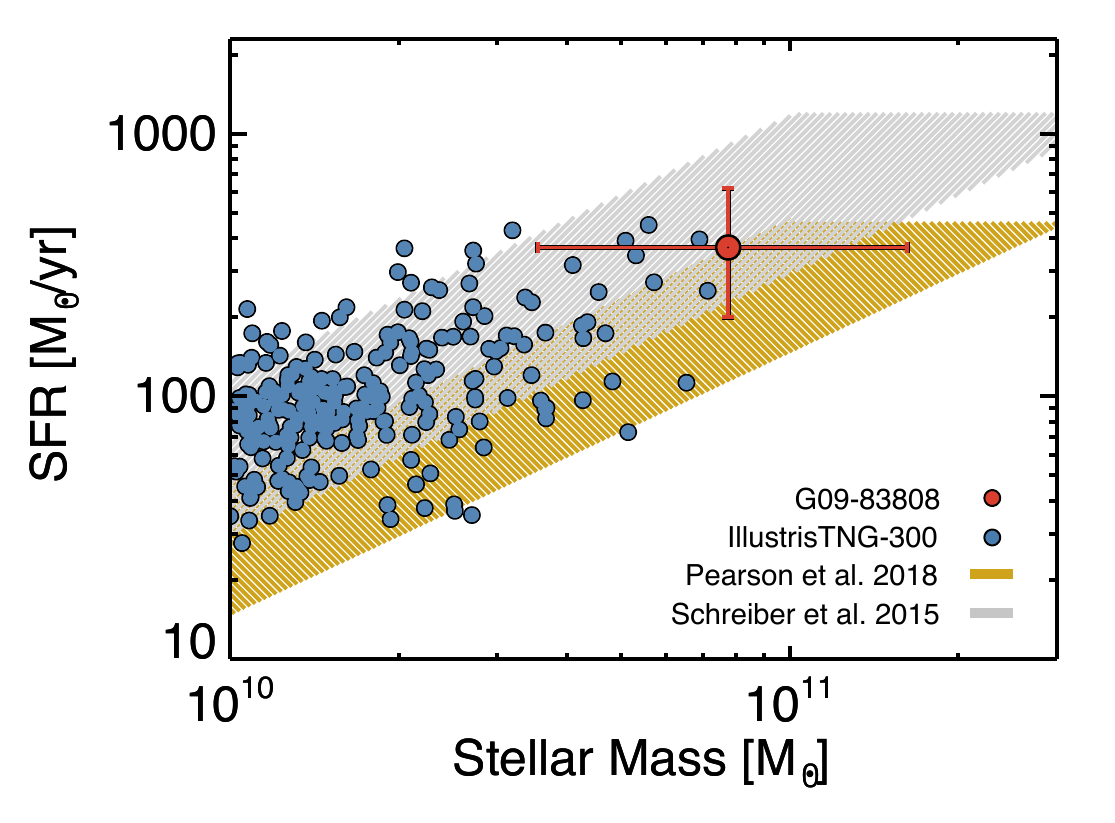}
\caption{{\it G09-83808 in the context of the main-sequence of star-forming galaxies.}
The best-fit SFR and stellar mass of G09-83808, and their associated uncertainties, are represented by the red solid circle. The gray and gold regions represents the main-sequence relationships of \citet{Schreiber2015a} and \citet{Pearson2018a}, respectively, extrapolated to $z=6$. 
For the sake of comparison we also include  $z\approx6$ star-forming galaxies from the IllustrisTNG-300 simulation (\citealt{Pillepich2018a}). Our target, G09-83808, lies on the high-mass end of the  main-sequence over a parameter space that overlaps with the the most massive star-forming galaxies from IllustrisTNG. 
\label{fig:main-sequence}}
\end{figure}

The $3.6$ and $4.5\,\mu\rm m$ {\it Spitzer}/IRAC observations were also used along with the FIR photometry to constrain the stellar mass of the galaxy through a spectral energy distribution (SED) fitting technique using the energy balance code {\sc magphys} (\citealt{daCunha2008a}). First, since the emission of G09-83808 is blended with that from the foreground $z=0.776$ lensing galaxy in the IRAC bands (see \citealt{Zavala2018a}), the light distribution of the foreground galaxy was modeled using {\sc galfit} (\citealt{Peng2002a}) and a S\'{e}rsic profile (\citealt{Sersic1963a}). Then, the emission from the foreground galaxy was subtracted from each image, and the photometry of the background source was measured from the {\sc galfit} generated residual images using  {\sc SExtractor} (\citealt{Bertin1996a}). Finally, combining the deblended {\it Spitzer} photometry, which probes the rest-frame optical stellar emission, with the FIR data, the best-fit SED for G09-83808 was derived, from which a stellar mass of $\rm M_\star=7.8^{+8.4}_{-4.2}\times10^{10}\,\rm M_\odot$ was inferred (after correcting for gravitational magnification). 

With these measurements in hand, 
we can place our target in the context of  
the main-sequence of star forming galaxies.

As it can be seen from Figure \ref{fig:main-sequence}, G09-83808 lies on the high-mass end of the main-sequence of star forming galaxies when compared to the extrapolated relation of \citet{Schreiber2015a}, or slightly above (with $\rm SFR/SFR_{\rm MS}\approx2-3$) if compared to the relation of \citet{Pearson2018a} or \citet{Khusanova2021} -- note that the cuts at high SFRs in the figure are imposed to represent the typical turnover at high masses (e.g. \citealt{Tomczak2016a}).


\begin{figure*}
\includegraphics[width=.35\textwidth]{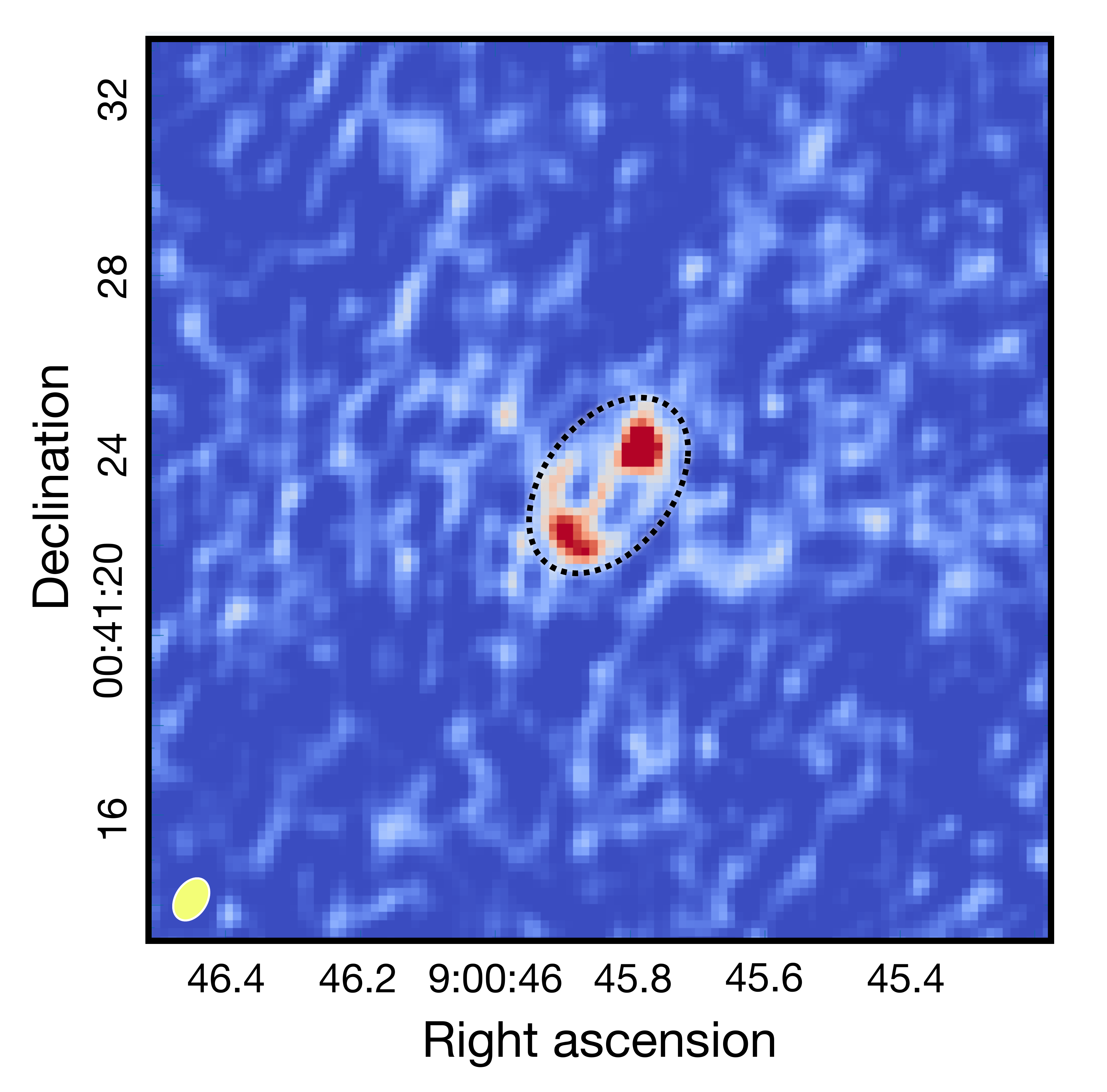}\hspace{0.4cm}\includegraphics[width=.63\textwidth]{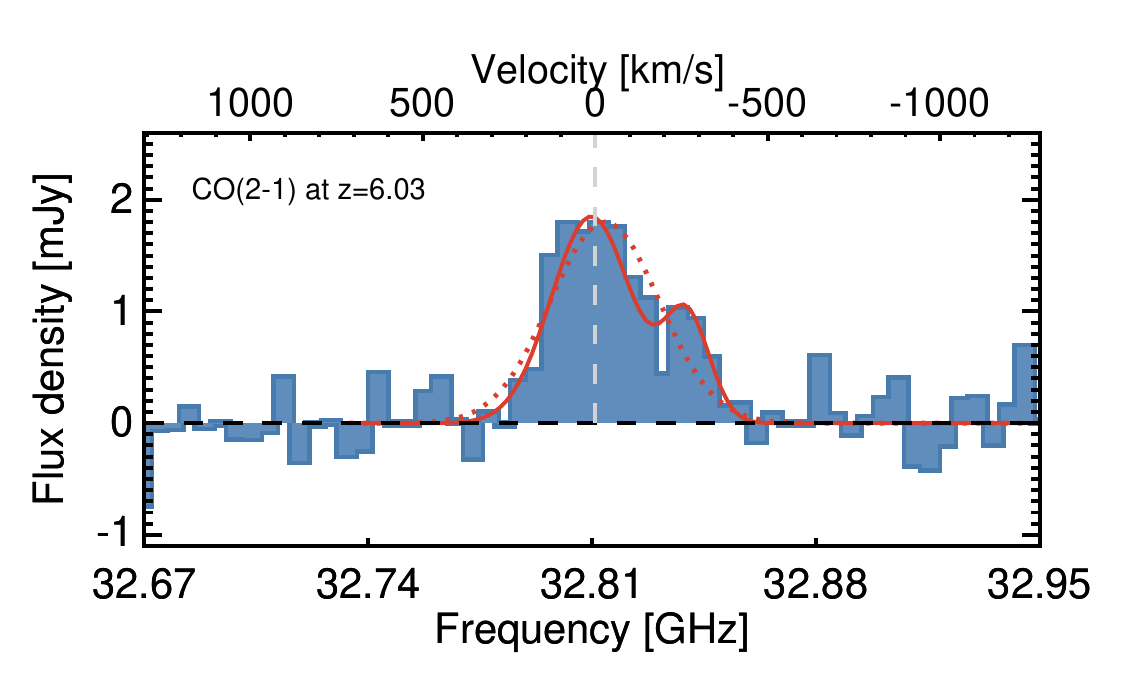}
\caption{{\it The $\rm CO(2\to1)$ transition from G09-83808. Left:}  Velocity-integrated intensity (moment-zero) map of the $\rm CO(2\to1)$ line over $\approx 800\,\rm km\,s^{-1}$ centered at $32.81\,$GHz. A double-arc structure is visible due to the gravitational lensing. The aperture used to extract the 1D spectrum is indicated with the dashed black line, while the beam-size of the observations is shown with the yellow ellipse in the bottom left.
{\it Right:} Extracted spectrum around the expected frequency of the line at $z=6.029$ (gray dashed line). The dotted red line represents the best-fit Gaussian function (with $\nu_{\rm obs}=32.810\pm0.002\,\rm GHz$ and $\rm FWHM=0.040\pm0.005\,\rm GHz$). The solid line represents the best-fit with two Gaussian components adopted to better describe the double-peak profile of the line, which is also noticeable in other transitions (see Appendix \ref{app:lines}).
\label{fig:line_detection}}
\end{figure*}

Since this $M_\star-\rm SFR$ relationship is still not well determined at $z\sim6$, we make use of results from simulations to further explore the place of our galaxy in the context of typical star-forming galaxies. To do this, we plot in Figure \ref{fig:main-sequence} the
$z\sim6$ star-forming galaxies from the IllustrisTNG project (\citealt{Pillepich2018a}). These galaxies 
are in relatively good agreement with the adopted main-sequence relationships, ruling out any significant bias in our comparison. As shown in this figure, the properties of our target are similar to those of the most massive galaxies in IllustrisTNG.  Therefore, we can conclude that G09-83808 probes the high-end of the main sequence of star-forming galaxies. 

This contrasts with the only other two $z>6$ galaxies discovered in blind (sub-)millimeter surveys, SPT0311-58 and HFLS3, which show SFRs of $\sim1,000\rm\,s$ of $\,\rm M_\odot\,\rm yr^{-1}$ and are considered to be extreme starburst galaxies (\citealt{Riechers2013a,Marrone2018a}).

Finally, we highlight that the metallicity of this galaxy has recently been constrained to be  $Z\approx0.5-0.7\,Z_\odot$ ($12+log(O/H)\approx8.34-8.54$) via the detection of [NII]$205\,\rm\mu m$ and [OIII]$88\,\rm\mu m$  in \citet{Tadaki2022a}. Interestingly, the mass-metallicity relation of \citet{Genzel2015a} predicts a value of $12+log(O/H)\approx8.48\pm0.07$, in very good agreement with \citet{Tadaki2022a}.

\subsection{VLA observations}

Observations were taken using the
Karl G. Jansky Very Large Array (VLA) in the C array configuration as part  of project 20A-386 (PI: J. Zavala). Three different executions were performed on 2020 June 10, 14, and 19 for a total integration time of 15\,h.

The WIDAR correlator setup was designed for simultaneous continuum and spectral line observations in the Ka-band, with mixed 3-bit and 8-bit samplers, resulting in a total bandwidth of 5.12\,GHz subdivided into forty 128\,MHz dual-polarization sub-bands with 1\,MHz channels.  
During each execution, the source $J1331+305$ (aka 3C-286) served as band-pass and absolute flux calibrator, while J$0909+0121$ was used as pointing source and complex gain calibrator. A few scans from some antennas were flagged before data calibration due to phase or amplitude issues or because they were affected by radio frequency interference (although they do not represent more than a few percent of all data). Data reduction and calibration were done using the VLA pipeline following the standard procedures. Then, the data from the three executions were combined during the imaging procedure, which was done using natural weighting of the visibilities in order to maximize the sensitivity at the expense of angular resolution (producing a synthesized beam size of $0.97\arcsec\times0.68\arcsec$, PA$=-30\,$deg). This results in an r.m.s noise level of $\approx60\,\mu\rm Jy/beam$ for a $\sim45\,\rm km\,s^{-1}$ channel width for the sub-band centered at the expected position of the $\rm CO(2\to1)$  line.

\section{Analysis and Results}
\subsection{The $\rm CO(2\to1)$  line and the lensing model}\label{secc:line_and_magnification}

Figure \ref{fig:line_detection} shows the $\rm CO(2\to1)$ line detection (moment-0 map and 1D extracted spectrum) from G09-83808. To measure the total flux density of the line we fit 
either one or two Gaussian profiles to the extracted spectrum (see Figure \ref{fig:line_detection}), and we measure it
directly from the clean moment-0 map. These methods give us a statistically consistent integrated line flux of $\mu S_{\rm CO(2\to1)}=0.6\pm0.1\rm\,Jy\,km\,s^{-1}$, which implies a line luminosity of $\mu L'_{\rm CO(2\to1)}=1.8\pm0.3\times10^{11} \,\rm K\,km\,s^{-1}\,pc^2$ (following \citealt{Solomon1992a}).

\begin{figure*}
\includegraphics[width=\textwidth]{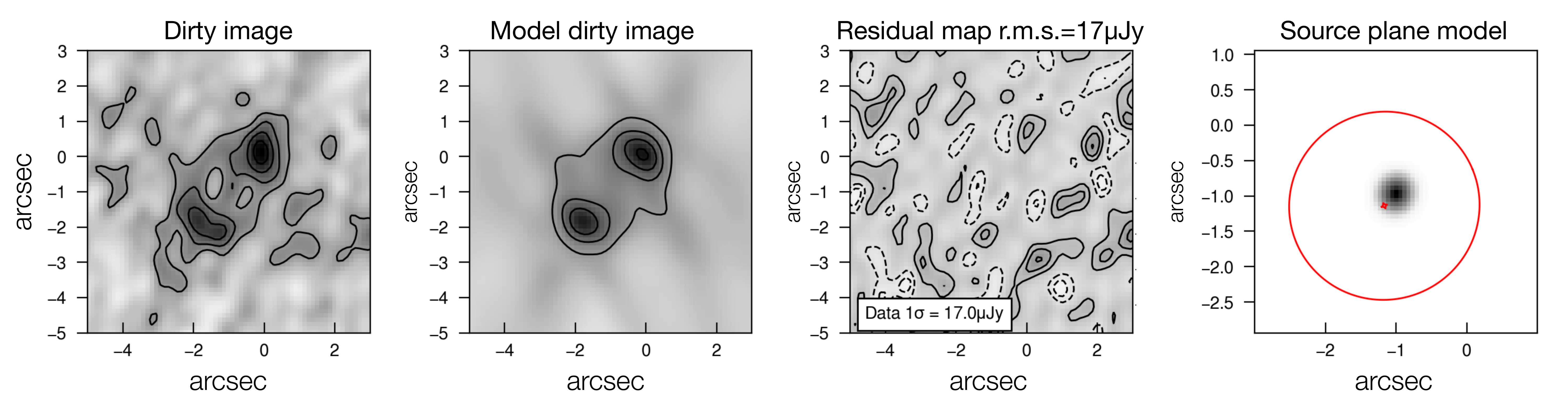}
\caption{{\it Lensing model.} From left to right, we show the dirty $\rm CO(2\to1)$ moment-zero map of G09-83808 obtained with the JVLA, the dirty image produced by the best-fit lensing model, the residual map (with $\pm1\sigma,\pm2\sigma,\pm3\sigma$ contours), and the map reconstruction in the source plane (with the lensing caustic represented by the red line). From this analysis, we infer a gravitational lensing magnification of $\mu=12\pm3$.
\label{fig:lensing_model}}
\end{figure*}


To measure the  gravitational amplification factor on the line, we use the lens modelling code {\sc visilens} (\citealt{Spilker2016a}), which directly models the visibilities in the {\it uv} plane. The modelling was done in a similar way as for the $890\,\mu$m dust continuum emission previously reported in \citet{Zavala2018a}, parameterizing the lens mass profile as a singular isothermal ellipsoid and the background source as a $n=1$  S\'ersic profile. 
While this modelling provides a good fit to the data (see Figure \ref{fig:lensing_model}), we acknowledge that the $\rm CO(2\to1)$ data alone cannot constrain the source profile. The assumed $n=1$ S\'ersic profile is however supported by the modelling of high-SNR, high-angular resolution observations of the dust continuum emission (\citealt{Zavala2018a,Tadaki2022a}.

After marginalizing over the different parameter space explored by the code, we derive a best-fit magnification factor of $\mu=12\pm3$  and a source-plane (intrinsic) size of $R_{\rm circ}=0.6\pm0.2\rm\,kpc$ ($0.11\pm0.03''$). These values are in  good agreement with the results presented in \citet{Zavala2018a}, who derived a magnification of $9.3\pm1.0$ and a size of  $R_{\rm circ}=0.6\pm0.1\rm\,kpc$ for the $890\,\rm\mu m$ dust continuum emission, and with \citet{Tadaki2022a} who found $\mu=8.4^{+0.7}_{-0.3}$ and $R_{\rm eff}\approx0.65\rm\,kpc$ for the  $1.5\,$mm continuum, despite using a different lensing code. 
This suggests that there is no significant differential magnification between the $\rm CO(2\to1)$ and the dust thermal emission and implies similar and co-spatial source plane areas.

Based on the $\rm CO(2\to1)$ magnification factor of $\mu=12\pm3$ described above, we derive a corrected $\rm CO(2\to1)$ line luminosity of $L'_{\rm CO(2\to1)}=(1.5\pm0.5)\times10^{10}\rm\,K\,km\,s^{-1}\,pc^2$.
\\




\subsubsection{The effect of the CMB}
The systematic increase of the temperature of the Cosmic Microwave Background (CMB) with increasing redshift ($T_{\rm CMB}\propto T^{z=0}_{\rm CMB}(1+z)$) affects the physical conditions and detectability of galaxies' dust and molecular gas emission. As detailed in \citet{daCunha2013a}, these effects are more important when the CMB temperature ($\approx 19\rm\,K$ at $z=6$) is close to the dust or gas temperature ($T_{\rm dust}$ or $T_{\rm kin}$, respectively).

\citet{Harrington2021a} and \citet{Jarugula2021a} have recently reported high kinetic temperatures ($T_{\rm kin}/T_{\rm d}\approx2-3$) for dusty star-forming galaxies.
If this holds at higher redshifts, it would imply that the CMB effects on the measured CO line fluxes are relatively minor (note that the CMB also affects the kinetic and dust temperatures; \citealt{daCunha2013a}). G09-83808, the source studied in this work, indeed shows a bright  $\rm CO(12\to11)$  detection as expected for a high kinetic temperature. Nevertheless, a detailed modelling of its SLED (including the  $\rm CO(2\to1)$,  $\rm CO(5\to4)$,  $\rm CO(6\to5)$, and $\rm CO(12\to11)$  transitions) suggests that the $\rm CO(2\to1)$ line is almost totally dominated by a relatively cold component with $T_{\rm kin}=55\pm16\,$K (Tsujita et al. submitted). 

Assuming this kinetic temperature and following \citet{daCunha2013a}, we infer a CMB correction factor of $1.14^{+0.04}_{-0.06}$ (note that we used the dust temperature as the minimum kinetic temperature in this calculation). The intrinsic $\rm CO(2\to1)$ flux density is then estimated to be $0.06\pm0.02\,\rm Jy\,km\,s^{-1}$, 
which translates into an intrinsic line luminosity  of $L'_{\rm CO(2\to1)}=(1.7\pm0.5)\times10^{10}\rm\,K\,km\,s^{-1}\,pc^2$. These values, which have been corrected by the gravitational amplification,  will be adopted for the rest of the analysis.


\subsection{Total molecular gas content}\label{secc:total_gas_mass}
The   $\rm CO(2\to1)$--to--$\rm CO(1\to)$  line ratios reported in the literature show a small scatter of less than a factor of 2 (e.g. \citealt{Harrington2021a}). This contrasts with the high dispersion in the line ratios between the $\rm CO(1\to0)$ ground transition and higher-J lines, which could differ by up to a factor of $\sim 6$ (see the CO SLED of DSFGs reported in \citealt{Casey2014a}; see also \citealt{Carilli2013a}). This implies that  the CO ground state luminosity, and thus the molecular gas mass, can be inferred with relatively little extrapolation from the 
$\rm CO(2\to1)$ detection.

Adopting a line brightness temperature ratio of $r_{2,1}=0.83\pm0.10$ (e.g. \citealt{Bothwell2013a,Carilli2013a,Genzel2015a,Harrington2021a}), we infer a $\rm CO(1\to0)$ line luminosity of $L'_{\rm CO(1\to0)}=(2.0\pm0.7)\times10^{10} \,\rm K\,km\,s^{-1}\,pc^2$
(after correcting for gravitational amplification).

As mentioned in Section \ref{sec:intro}, the total molecular gas mass can be directly inferred from the $\rm CO(1\to0)$ line luminosity via the CO-to-H$_2$ conversion factor, $\alpha_{\rm CO}$. However, this conversion factor suffers from large uncertainties and might depend on several physical parameters of a galaxy's ISM (e.g. \citealt{Bolatto2013a}). 
A value of $\alpha_{\rm CO}=1.0\,\rm M_\odot\,K\,km\,s^{-1}\,pc^2$ is typically 
adopted for nearby nuclear starburst galaxies\footnote{Historically, the ULIRG value was adopted to be
$\alpha_{\rm CO}=0.8\,\rm M_\odot\,K\,km\,s^{-1}\,pc^2$, nevertheless, 
a value of $\alpha_{\rm CO}=1\,\rm M_\odot\,K\,km\,s^{-1}\,pc^2$ aligns better with the more recent review by \citealt{Bolatto2013a}.} (i.e. ULIRGs; \citealt{Downes1998a}), while a higher value of $\alpha_{\rm CO}=4.6\,\rm M_\odot\,K\,km\,s^{-1}\,pc^2$ is thought to be more appropriate for more moderate galaxies (including the Milky Way; see review by  \citealt{Bolatto2013a}). These two extreme values would bracket the molecular gas of G09-83808 between $\sim2\times10^{10}\,\rm M_\odot$ and $\sim9\times10^{10}\,\rm M_\odot$.

Very high $\alpha_{\rm CO}$ could, however, lead to a molecular gas mass estimate larger than the dynamical mass (e.g. \citealt{Bryant1999a}). Such dynamical information can thus be used to constrain the CO-to-H$_2$ conversion factor in individual systems (e.g. \citealt{Bothwell2010a,Engel2010a}). In this work we exploit the dynamical information encoded in the $\rm CO(2\to1)$ emission line that traces the cold gas reservoir to restrict the CO-to-H$_2$ conversion factor of our target. 

The dynamical mass is estimated by:
\begin{equation}
    {\rm sin^2}(i)\,\rm M_{\rm dyn} =  \frac{5R_{\rm e}\sigma^2}{G},
\end{equation}
where $R_{\rm e}$ is the effective circularized radius, $\sigma$ is the velocity dispersion, and G the gravitational constant (e.g. \citealt{Casey2019a}). To correct for the unknown inclination, $i$, we multiply by a factor of $3/2$ (the reciprocal of the expectation value of ${\rm sin^2}(i)$).  Based on this equation, we infer a dynamical mass of $\rm M_{\rm dyn} = (2.6^{+2.4}_{-1.5})\times10^{10}\,\rm M_\odot$ (note that a similar value is obtained if we use instead the isotropic viral estimator equation, e.g. \citealt{Engel2010a}). This places an upper limit on the conversion factor of $\alpha_{\rm CO}<2.5\,\rm M_\odot\,K\,km\,s^{-1}\,pc^2$ (taking into account $1\sigma$ uncertainties), which justifies the adoption of $\alpha_{\rm CO}=1.0\,\rm M_\odot\,K\,km\,s^{-1}\,pc^2$, as typically measured for ULIRGs (\citealt{Bolatto2013a}).
The  total de-lensed molecular gas mass of G09-83808 is then estimated to be $\rm M_{\rm H_2}=(2.0\pm0.7)\times10^{10}\,\rm M_\odot$.

\subsubsection{A Test on the dust continuum method}
\citealt{Scoville2016a} presented a relationship between the specific luminosity at rest-frame $850\,\mu\rm m$, $L_\nu{(850\mu\rm m)}$, and the CO line luminosity, $L'_{\rm CO(1\to0)}$. Such a relation serves as the base for an empirical calibration aimed at inferring galaxies' molecular gas content via dust continuum observations that probe the Rayleigh-Jeans tail of dust black-body emission. While this time-efficient approach has been revolutionary for the measurement of ISM masses --- because dust continuum is observationally cheaper than spectral observations of CO transitions, its calibration and validity have not been tested beyond $z\sim4$ (even when studies using this method extends up to $z\sim6$; e.g. \citealt{Liu2019a}).

Thanks to the $\rm CO(2\to1)$ detection 
in G09-83808, here we test whether or not the $L_\nu{(850\mu\rm m)}$-$L'_{\rm CO(1\to0)}$ calibration holds at $z=6$ for this system. Note that we decide to test this relation rather than the final molecular gas recipe of \citealt{Scoville2016a} to avoid possible differences in the $\alpha_{\rm CO}$ assumptions.


To mimic the typical procedure adopted in the literature, we calculate the $850\,\mu\rm m$ specific luminosity from one single data-point close to the Rayleigh-Jeans tail (in this case, the gravitationally-corrected $1.1\,\rm mm$ flux density of $S_{\rm 1.1mm}=2.2\pm0.3$ reported in \citealt{Zavala2018a}). This calculation assumes a dust emissivity index of $\beta=1.8$ and a mass-weighted dust temperature of $25\,\rm K$ (see details in \citealt{Scoville2016a}). 

The \citealt{Scoville2016a} calibration predicts a $\rm CO(1\to0)$ line luminosity of $(4.6\pm1.0)\times10^{10}\,\rm K\,km\,s^{-1}\,pc^2$. This is a factor of $2.3\pm0.9$ above the $\rm CO(2\to1)$-derived value of $(2.0\pm 0.7)\times10^{10}\,\rm K\,km\,s^{-1}\,pc^2$.

As mentioned above, the  alternative approach typically adopted in the literature to estimate the $\rm CO(1\to0)$ line luminosity from high-J transitions might result in larger discrepancies than the one observed here.  Therefore, although larger samples are imperative to derive any firm conclusions,
this result supports the empirical approach of using single band continuum observations as a proxy for molecular gas content up to $z\sim6$ (modulo an appropriate $\alpha_{\rm CO}$)\footnote{The reader should keep in mind, though, that the final  \citeauthor{Scoville2016a} equation to translate from dust continuum to interstellar medium mass should be modified appropriately if a different $\alpha_{\rm CO}$ value wants to be adopted.}. We highlight, though, that the CMB effects could be more pronounced in galaxies at higher redshift and/or with  colder temperatures.

An important corollary that can be inferred from this, is that the gas-to-dust abundance ratio at $z\sim6$ is similar to that found in lower redshift galaxies, at least for this massive (and relatively metal-rich) galaxy. 
This is in line with some recent simulations which predict a modest redshift evolution on the gas-to-dust ratio (e.g. \citealt{Li2019a,Popping2022a}).

\subsection{Gas depletion timescale, gas-to-dust ratio, and gas fraction}\label{secc:gas_relations}

\begin{figure*}\hspace{-0.6cm}
\includegraphics[width=0.53\textwidth]{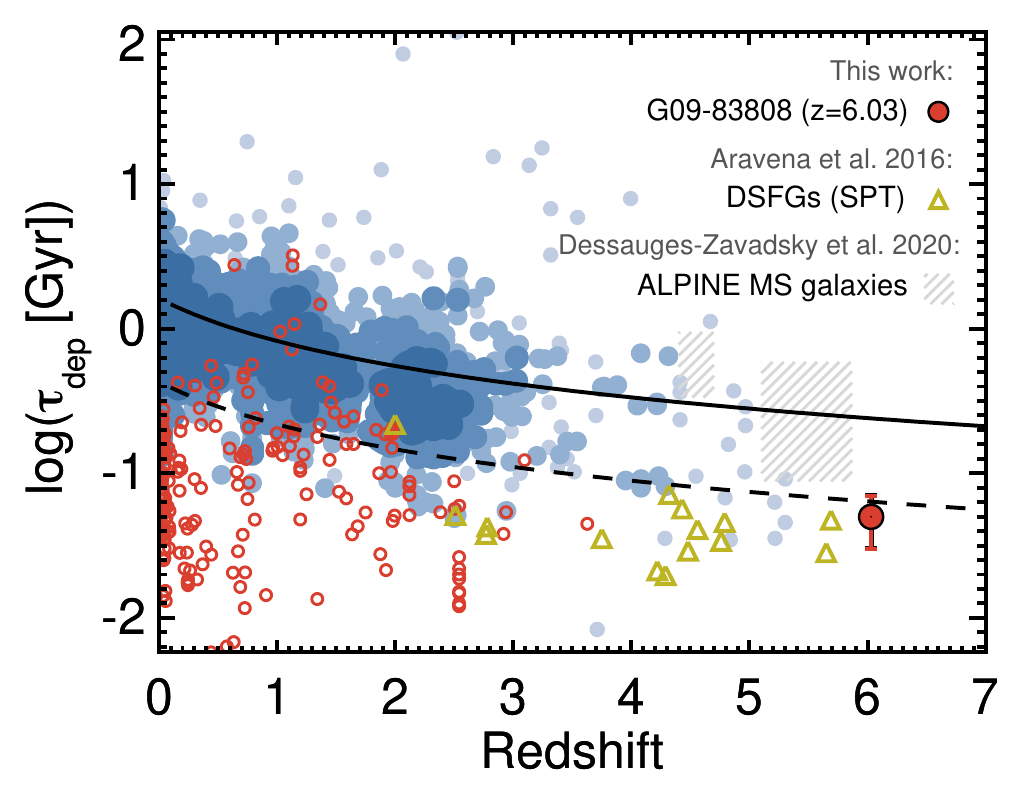}\hspace{-0.2cm}
\includegraphics[width=0.53\textwidth]{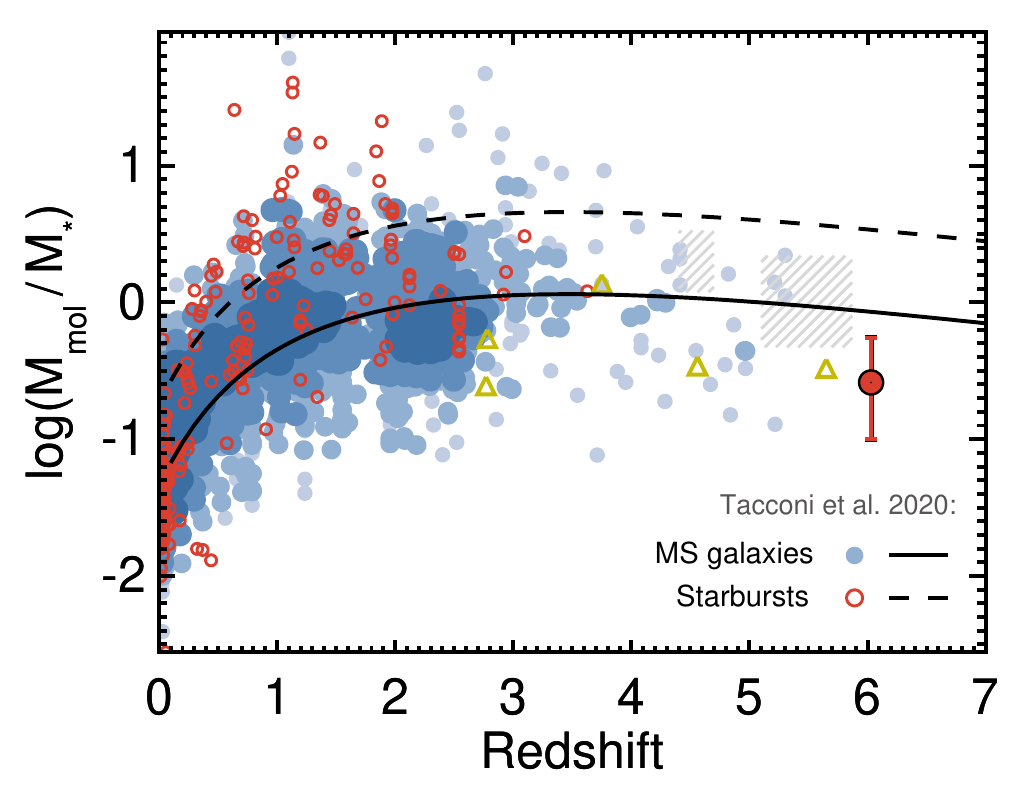}
\caption{{\it The evolution of the gas depletion time and gas fraction.} {\it Left:} The derived depletion timescale of G09-83808, represented by the red solid circle, is shown against the field scaling relation derived by \citet{Tacconi2020a} and their compiled sample (see \S3.3 for more details). Blue circles represent main-sequence star-forming galaxies ($log(\rm SFR/SFR_{\rm MS})<1$) in such compilation, where the darkness and size represent a proxy for the number density. Open red circles represent their starburst galaxies ($log\rm (SFR/SFR_{\rm MS})\ge1$) with gas masses scaled down to adjust for $\alpha_{\rm CO}$ (see \S\ref{secc:gas_relations}). The solid and dashed lines are the best-fit relations for main-sequence ($log(\rm SFR/SFR_{\rm MS})=0$)  and starburst galaxies ($log(\rm SFR/SFR_{\rm MS})\approx1.2$) , respectively, extrapolated to $z\sim7$. We also include the sample of dusty star-forming galaxies of \citet{Aravena2016a} selected with the SPT (yellow triangles) and highlight the locus occupied by the ALPINE main-sequence galaxies (\citealt{Dessauges-Zavadsky2020a}; gray regions). While the scaling relation successfully captures the behavior towards shorter depletion timescales for starbursts, a significant fraction of them scatter below the relationship, suggesting even shorter depletion times (or higher star formation efficiencies).
{\it Right:} Analog to the left panel (with same symbols and colors) but for the gas-to-stellar mass ratio. For the SPT sample, we only plot those sources with stellar mass measurements in \citealt{Ma2015a}. While some $z\approx1-2$ starbursts have high gas fractions consistent with the best-fit scaling relation, G09-83808, the SPT galaxies, and most of the 
local ($z\sim0$) ULIRGs do not show enhanced gas fractions with respect to main-sequence galaxies.
\label{fig:scaling_relations}}
\end{figure*}

With a gas mass measurement in hand, we can now explore other parameters such as the gas-to-dust ratio and gas fraction, and the gas depletion timescale, and compare them to those measured in other galaxies. In addition, we can test whether or not the scaling relations found for lower redshift systems still hold at $z\sim6$. This is the main goal of this section.

For this comparison we use the recent compilation by \citet{Tacconi2020a} of $\sim2,000$ galaxies with gas measurements and their derived relationships, which are built upon previous significant efforts (e.g. \citealt{Genzel2015a,Scoville2017a,Tacconi2018a}). Although this collection includes galaxies up to $z\sim5$, the low number of galaxies at high redshifts limits the analysis and the derived relationships to $z\lesssim4$. Hence, our observations allow us to investigate the evolution of these parameters within an unexplored redshift range and test the validity of the extrapolated scaling relations at $z\sim6$. 

It is important to highlight that \citet{Tacconi2020a} used a CO conversion factor of $\alpha_{\rm CO}=4.36\,\rm M_\odot\,K\,km\,s^{-1}\,pc^2$ and a gas-to-dust ratio of $\delta_{\rm GDR}=67$ to re-calibrate all the gas mass measurements from the literature homogeneously. This is justified since their analysis is focused on galaxies around the main-sequence ($\Delta\rm MS\equiv log(\rm SFR/SFR_{\rm MS}):[-1,1]$), for which little to no systematic variation of these parameters have been found (see detailed discussion in \citealt{Genzel2015a,Scoville2017a,Tacconi2020a}).
Nevertheless, this has been shown not to be the case for extreme outliers with log$(\rm SFR/SFR_{\rm MS})>1$ (which represent only $\sim10\,$\% of their sample). Therefore, for those extreme galaxies with gas mass estimates based on $\alpha_{\rm CO}=4.36\,\rm M_\odot\,K\,km\,s^{-1}\,pc^2$, we have scaled the gas masses down by a factor of 4.36, and propagated this to other measurements, to reflect our assumption of $\alpha_{\rm CO}=1\,\rm M_\odot\,K\,km\,s^{-1}\,pc^2$. The appropriateness and the implications of this choice are discussed below.

\subsubsection{Gas-to-dust ratio}

To calculate the gas-to-dust ratio, we adopt the amplification-corrected gas mass reported above in \S\ref{secc:total_gas_mass} and the demagnified dust mass of $\rm M_{\rm d}=(1.9\pm0.4)\times10^8\,\rm M_\odot$ reported by \citet{Zavala2018a}. Using these values, we estimate a gas-to-dust ratio of 
$\delta_{\rm GDR} =105\pm40$, which is in excellent  agreement with the widely adopted ratio of 100:1 for massive galaxies with close to solar metallicities (e.g. \citealt{Leroy2011a,Remy-Ruyer2014a}) and consistent within $1\sigma$ with the value of $\sim67$ adopted in the recent review by \citet{Tacconi2020a}.

Given that G09-83808 was shown to have a near-solar-metallicity ISM (\citealt{Tadaki2022a}), this result supports the conclusion by \citet{Remy-Ruyer2014a} who claimed that the metallicity is the main physical property driving the gas-to-dust ratio (see also \citealt{Li2019a,Popping2022a}), and it suggests that this relation does not significantly evolve with redshift (c.f. \citealt{Saintonge2013a}).

In addition, this supports our assumption of $\alpha_{\rm CO}=1\,\rm M_\odot\,K\,km\,s^{-1}\,pc^2$ for the gas estimation. Adopting a higher value (as the typical Milky Way value) would not only result in a gas mass in excess of the dynamical mass, but also in an relatively high gas-to-dust ratio of around $\delta_{\rm GDR}\sim500$.

\subsubsection{Depletion time}\label{secc:gas_depl}

The depletion timescale, which expresses the time in which the molecular gas reservoir would be consumed at the current SFR (in the absence of inflows or outflows of molecular gas), is calculated as $\tau_{\rm depl}=\rm M_{\rm mol}/\rm SFR$ (note that the inverse of this quantity is usually referred to as the star formation efficiency, $\rm SFE=1/\tau_{\rm depl}$). For this calculation we assume the molecular gas mass derived above (\S\ref{secc:total_gas_mass}) and the SFR of $380\pm50\rm\,M_\odot\,yr^{-1}$ reported by \citet{Zavala2018a}. Combining these measurements we infer a gas depletion time of $\tau_{\rm depl}=50\pm20\rm\,Myr$ for G09-83808.

In Figure \ref{fig:scaling_relations} we compare the gas depletion time of G09-83808 with the  large compilation of \citet{Tacconi2020a} and their scaling relation. \citet{Tacconi2020a} found that the 
integrated depletion time scale depends mainly on redshift and offset from the main-sequence ($\tau_{\rm depl}\propto\rm (1+z)^{-1}\times\Delta{\rm MS}^{-0.5}$). This trend has been confirmed to extend up to $z\sim5.5$ thanks to the ALPINE survey (\citealt{Dessauges-Zavadsky2020a}). The inferred depletion time of our target is indeed consistent with this trend of decreasing depletion time with redshift, although significantly shorter than the expected relation for main-sequence galaxies. Surprisingly, its  depletion of $\tau_{\rm depl}=50\pm20\rm\,Myr$  is in  better  agreement with the extrapolated relation for starburst galaxies (see Figure \ref{fig:scaling_relations}), despite being located on the main-sequence of star-forming galaxies (as discussed in \S\ref{sec:target}).

To extend our comparison to other high-redshift dusty star-forming galaxies, we also include in Figure \ref{fig:scaling_relations} the seventeen sources reported in \citet{Aravena2016a}, which cover a redshift range of $z\approx2.5-5.5$ with a median SFR of $\sim1,000\,\rm M_\odot\, yr^{-1}$. All these galaxies, including G09-83808 and the most extreme galaxies in the \citet{Tacconi2020a}  compilation (with $log(\rm SFR/SFR_{\rm MS})>1$; open red circles in the figure), are in better agreement with the \citeauthor{Tacconi2020a} relation for starburst galaxies (see dashed black line on the left panel of Figure \ref{fig:scaling_relations}).

By looking carefully at Figure \ref{fig:scaling_relations}, one can realize, though, that a significant fraction of these extreme galaxies lie even below this starburst relationship. This is true for almost all of the $z\sim0$ ULIRGs and the SPT galaxies, which suggests that all these extreme galaxies might have even higher star formation efficiencies.
\citet{Aravena2016a} concluded that most of the galaxies in their sample are indeed experiencing a starburst phase likely triggered by major mergers. This triggering mechanism could thus explain the shorter depletion times (i.e. higher star formation efficiencies) of these galaxies compared to galaxies around the main-sequence. Similarly, local ULIRGs and other outlier objects are also known to be almost invariably mergers  with very compact nuclei (\citealt{Sanders1991a}), which supports the existence of a  {\it starburst} mode of star formation (e.g. \citealt{Sargent2012a,Silverman2015}).

The double-peak profile in the lines of G09-83808 (see Figure \ref{fig:line_detection} and Appendix \ref{app:lines})  might indeed be indicative of a merger activity, as in the case of the core of Arp 220 (\citealt{Downes1998a}) and other DSFGs (e.g. \citealt{Bothwell2013a}).
Nevertheless, Tsujita et al. (submitted) showed, based on a source-plane reconstruction of the [OIII]88$\mu\rm m$ and [NII]205$\mu\rm m$ emission lines across different velocity channels, that this galaxy is well-modelled by a rotating disk with some compact star-forming clumps. It is thus likely that the high star formation efficiency of this galaxy and its starburst-like properties are the consequence of gas compression induced by its compact size (e.g. \citealt{Elbaz2011a,Gomez-Guijarro2022}).

\subsubsection{Gas fraction}\label{secc:gas_fracc}
Here, we explore the gas fraction (in terms of the gas-to-stellar mass ratio) of our target and place it in the context of previous studies.

The redshift evolution of this quantity, and its dependency with other parameters, was also studied and parameterized in the recent review by \citet{Tacconi2020a}. They concluded that the redshift evolution of $\mu_{\rm mol}=\rm M_{\rm mol}/M_\star$ is better described by a quadratic function, with a steep rise between $z=0$ and 
$z\sim2-3$, the redshift at which the galaxies show the largest molecular fractions (at fixed $\rm SFR/SFR_{\rm MS}$), and with a subtle turnover at higher redshifts (Figure \ref{fig:scaling_relations}).  This flattening (or slight turnover) at high redshift is also supported by the recent results of \citet{Dessauges-Zavadsky2020a}, who measured an average gas fraction of $\mu_{\rm mol}\approx60\%$ over $z\approx4.5-5.5$ (although with a relatively large scatter driven likely by galaxies' stellar masses, as can bee seen in the figure). 
To lesser degree, this parameter also depends on the vertical location of a galaxy along the main-sequence plane as seen in Figure \ref{fig:scaling_relations}. Such a relation implies that the increase in galaxies' SFRs with redshift (at fixed $\rm SFR/SFR_{\rm MS}$) and with offset from the main sequence (at fixed redshift) is due primarily to increased gas content, and secondarily to an increased efficiency for converting gas to stars -- at least for those `normal' galaxies around the main-sequence (the focus of the \citealt{Tacconi2020a} study).

Figure \ref{fig:scaling_relations} reveals that some $z\sim1-2$ 
starbursts do also have high gas fractions, suggesting that their high SFRs could also be the result of having a large gas mass reservoir, as seen for modestly star-forming galaxies. Nevertheless, most of the local ULIRGs do not show  enhanced gas fractions. The same is true for the few SPT-selected galaxies with stellar mass measurements (\citealt{Ma2015a}), whose  gas-to-stellar mass ratios are in agreement with `normal' star-forming galaxies (see Figure  \ref{fig:scaling_relations}).
Starburst galaxies could thus have a large range of gas fractions, implying that their high SFRs are not always caused by their high gas masses but also by other mechanisms that enhance their star formation efficiencies, as discussed before in the literature (e.g. \citealt{Scoville2017a}).
 
The galaxy studied in this work has, interestingly, a relatively low gas fraction of $\mu_{\rm mol}=0.26^{+0.29}_{-0.16}$
(see red solid circle in Figure \ref{fig:scaling_relations}), comparable to those of the SPT sample. This supports the scenario discussed above: G09-83808 shows a {\it starburst} mode of star formation with a high star formation efficiency driven likely by its internal properties such as gravitational instabilities or gas compression.

\subsection{An alternative scenario with a Milky Way-like CO-to-H$_2$ conversion factor }

While a low ULIRG-like CO-to-H$_2$ conversion factor ($\alpha_{\rm CO}\sim1\,\rm M_\odot\,K\,km\,s^{-1}\,pc^2$) is favored by the current analysis thanks to the limits placed by the dynamical mass, we cannot totally discard the possibility of a higher value, particularly if the system is experiencing a merger
(since the dynamical mass assumes a virialized system).
The source-plane reconstruction of this galaxy is well-modelled with a disk profile though (see \S\ref{secc:line_and_magnification} and \citealt{Tadaki2022a} who used 
higher angular resolution observations with a physical spatial resolution below $1\,\rm kpc$ in the source plane). In addition, the
[OIII]88$\mu\rm m$ and [NII]205$\mu\rm m$ emission lines show evidence of a monotonic velocity gradient consistent with a rotating disk (Tsujita et al. submitted). This backs the applicability of a dynamical estimator for the total gas of G09-838083 and the inferred upper limit  on the conversion factor of $\alpha_{\rm CO}<2.5\,\rm M_\odot\,K\,km\,s^{-1}\,pc^2$.

Assuming a larger value of $\alpha_{\rm CO}$
would not only imply a molecular gas mass exceeding the dynamical mass but also a 
gas-to-dust ratio of  $\sim500$ (if assuming the Milky Way conversion factor reported in \citealt{Bolatto2013a}); at odds with the typical values measured for massive and metal-rich star-forming galaxies.
A galaxy with $\delta_{\rm GDR}=500$ would be expected to have a relatively low metallicity in the range of  $Z\approx0.15-0.35Z_\odot$ (according to the $\delta_{\rm GDR}$-metallicity relations of \citealt{Remy-Ruyer2014a,Tacconi2018a,Boogaard2021a}). This contrasts with the metallicity of G09-838083 which was constrained to  $Z\approx0.5-0.7\,Z_\odot$  by \citet{Tadaki2022a}. Therefore, given the current constraints, the possibility of G09-83808 having a large 
Milky Way-like CO-to-H$_2$ conversion factor seems unlikely.

\section{Discussion and conclusions}
Taking advantage of gravitational amplification, here we present the first detection of $\rm CO(2\to1)$ in a galaxy lying on the main-sequence of the $\rm M_\star-SFR$ plane and the characterization of its cold ISM and 
star formation.

While G09-83808 lies around the scatter of the main-sequence of star-forming galaxies at $z\sim6$ (see Figure \ref{fig:main-sequence}), it shows properties that resemble those from local ULIRGs and other starbursts, including a compact (dust and gas) size of $R_{\rm circ}=0.6\pm0.2\,$kpc (\S\ref{secc:line_and_magnification}), a relatively high dust temperature ($T_{\rm d}=49\pm3\rm\,K$; \citealt{Zavala2018a}), a low $\alpha_{\rm CO}$ of $<2.5\,\rm M_\odot\,K\,km\,s^{-1}\,pc^2$ (\S\ref{secc:total_gas_mass}),  and a short depletion time ($\tau_{\rm dep}=50\pm20\,\rm Myr$; \S\ref{secc:gas_depl}). All this suggests that galaxies with starburst-like ISM properties and star formation modes could also be hidden within the main-sequence population (see also \citealt{Gomez-Guijarro2022}).


\begin{figure}
\hspace{-0.5cm}\includegraphics[width=0.53\textwidth]{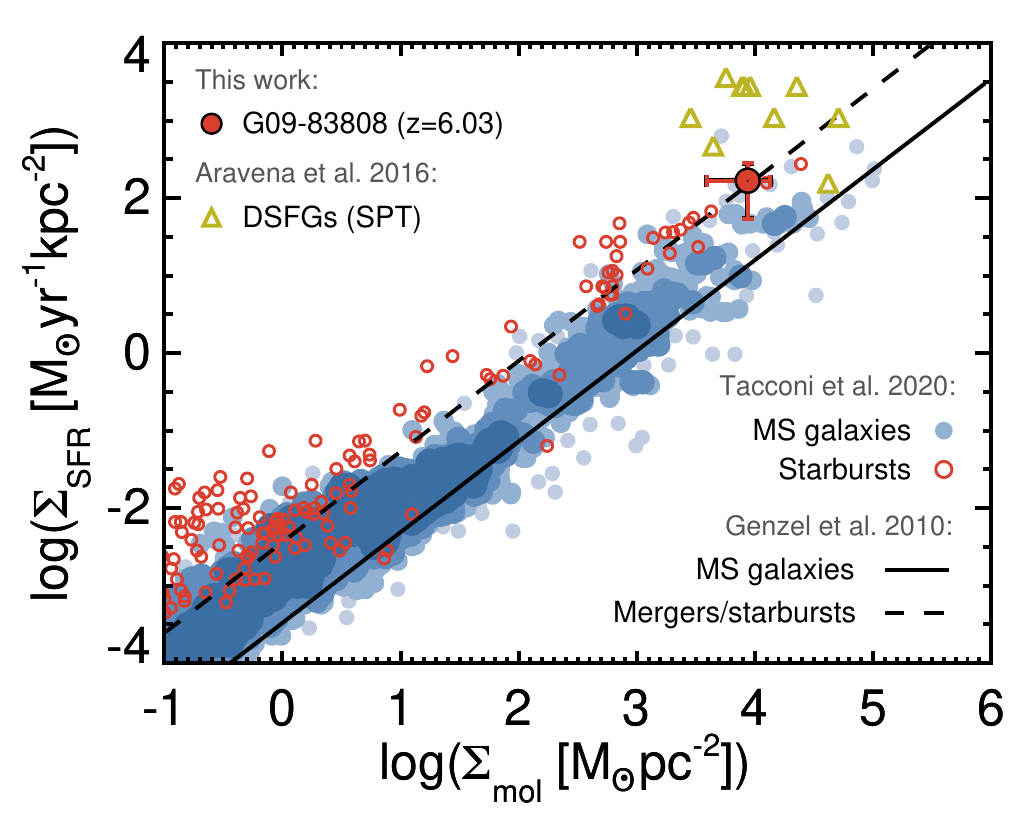}
\caption{{\it The Kennicutt-Schmidt ($\Sigma_{\rm gas}-\Sigma_{\rm SFR}$) relation.} 
The molecular and star formation rate surface density
of our $z=6$ target, G09-83808, is shown in comparison to other galaxies (following the same color code as in Figure \ref{fig:scaling_relations}). Note that for the \citet{Tacconi2020a} compilation we use the reported optical effective radius as a proxy for the size of the star forming regions and the molecular reservoirs (and our scaled gas masses for the starburst galaxies; see \S\ref{secc:gas_relations}). For the SPT sample we use the 
dust-based sizes from \citealt{Spilker2016a}.
Despite the assumptions on the sizes, all these starburst galaxies, including G09-83808, lie on the \citet{Genzel2010a} relation for merger/starburst galaxies (dashed black line), while the main-sequence galaxies lie below in better agreement with the relation for `normal' star-forming galaxies (black solid line).
This diagram reveals the starburst nature of our target, which is not captured by the main-sequence plane.
\label{fig:resolved_KS}}
\end{figure}


The starburst-like nature of G09-83808 is further revealed by the Kennicutt-Schmidt relation (Figure \ref{fig:resolved_KS}). This galaxy shows a high star formation rate surface density ($\Sigma_{\rm SFR}=170\pm110\rm M_\odot\,yr^{-1}\,kpc^2$) in agreement with what is expected for merger/starburst galaxies, according to the relation of  \citealt{Genzel2010a} (see dashed black line in the figure). Previous studies have found that galaxies with high $\Sigma_{\rm SFR}$ show shorter depletion timescales, smaller gas fractions, warmer dust temperatures, and high excitation conditions (e.g.  \citealt{Narayanan2014a,Franco2020a,Burnham2021a,Puglisi2021a}) in line with the properties of our target. We thus hypothesize that the star formation rate surface density can be a better indicator of starburst-like galaxies than the main-sequence offset (or at least, alternative to).

Despite its starburst-like nature, this high-redshift ULIRG analog, shows a relatively low gas fraction of
$\rm M_{\rm gas}/M_\star=0.29^{+0.33}_{-0.17}$ (see Figure \ref{fig:scaling_relations} and Section \ref{secc:gas_fracc}), implying that its starburst episode is not produced by an increased gas supply, but rather by a physical triggering mechanism that enhances its star formation efficiency. Typical mechanisms cited in the literature to explain this starburst mode of star formation include major/minor mergers, compactness, and disk instabilities. 
While higher angular resolutions are required
to precisely determine the causes of the high star formation efficiency in G09-83808, we attribute it to its internal properties, such as compactness and/or gravitational instabilities which facilitate gas compression (see further discussion in Tsujita et al. submitted, who found evidence of star-forming clumps within the rotating disk of this system). This is in line with recent results from \citet{Gomez-Guijarro2022}, who found that the most compact galaxies show the shortest depletion time and a starburst-like modes of star formation.


Without further significant gas accretion, we estimate,  following \citet{Casey2021a}, that G09-83808 will finish its star formation episode at $z\approx5.6-5.7$ with a final stellar mass on the order of $1\times10^{11}-3\times10^{11}\,\rm M_\odot$. This is well aligned with the derived masses of massive quiescent galaxies discovered at high redshifts (e.g. \citealt{Tanaka2019a,Marsan2020a,Valentino2020a,Santini2021a,Stevans2021a}), which also show compact sizes similar to the size of our target. This hints at a progenitor-descendant scenario between these populations of galaxies, and support other results from the literature suggesting that compact galaxies with low gas fractions and short depletion times are likely in the latest stages of a starburst episode and on the verge of quenching (e.g. \citealt{vanDokkum2015,Spilker2016b,Puglisi2021a,Gomez-Guijarro2022,Ikarashi2022}).

The determination of the molecular gas mass via direct detection of a low-J CO line transition also allows us to test whether or not other widely used methods to infer gas masses are also valid at $z=6$.
We found that the relationship between the rest-frame $850\rm\,\mu m$ specific luminosity and CO line luminosity ($L_\nu(850_{\rm \mu m})-L'_{\rm CO(1\to0)}$), on which the \citealt{Scoville2016a} calibration for estimating ISM masses from dust continuum emission is built on, satisfactorily predicts the $\rm CO(1\to0)$ line luminosity for this system within a factor of $\sim2$. 
Therefore,  single band continuum observations seem to be a promising way to constrain galaxies' gas masses even at this early epoch (modulo the assumption of a similar $\alpha_{\rm CO}$). 
A corollary that could follow this result is that the gas-to-dust ratio, $\delta_{\rm GDR}$, in massive (metal-rich) star-forming galaxies does not significantly evolve with redshift, as predicted by recent simulations (e.g. \citealt{Li2019a,Popping2022a}).  This is independently  confirmed for our target  using  the gas and dust estimates, from which we derive a $\delta_{\rm GDR}=105\pm40$. This value is also in relatively good agreement with the expected $\delta_{\rm GDR}$
given its high metallicity of $Z\approx0.5-0.7\,Z_\odot$ (if assuming a linear correlation between gas-to-dust ratio and metallicity, as those reported in \citealt{Tacconi2018a} and \citealt{Boogaard2021a}).


This experiment would not have been possible without the gravitational amplification effect on this galaxy, as of order $\sim1000\,$h on the VLA would have been required. In the near future, the ALMA Band-1 receiver (\citealt{Huang2016a}) might be able to detect $\rm CO(2\to1)$ in similar unlensed galaxies with observations on the order of a few $10\rm\,s\,$ of hours although limited to $z\lesssim5.5$.
The  detection  of  cold  molecular gas  in  ‘normal’, unlensed  star-forming galaxies at higher redshifts is thus prohibitive with current facilities and requires a ten-fold improvement in sensitivity. This stresses the necessity of new generation facilities such as the next-generation VLA (ngVLA, \citealt{Carilli2015a}) to directly pin down the molecular gas content of galaxies within the epoch of reionization and to calibrate alternative tracers.

Given that the gravitational amplification factor in this galaxy is similar to the expected increase in sensitivity of the ngVLA, this work can be seen as a preview of the unique expected capabilities of the ngVLA to study the cold ISM in galaxies within the first billion years of the Universe. At the same time, this detection solves previous concerns about the detectability of such emission lines due to the
increasing temperature of the CMB, anticipating the success of future radio facilities, although we note that the CMB impact could be higher in more extended sources with lower gas kinetic temperatures.







\begin{acknowledgments}

We thank Mark Lacy, Drew Medlin, and Aaron Lawson for their support with data reduction and the organizers of the NRAO 8th VLA Data Reduction Workshop. We also thank Kevin Harrington, Ryota Ikeda, Ikki Mitsuhashi, and Eric Murphy for fruitful discussions during the preparation of this manuscript and Gerg\"o Popping for facilitating the catalogs of simulated galaxies. Finally, we thank the anonymous reviewer for a constructive report and valuable suggestions.

The National Radio Astronomy Observatory is a facility of the National Science Foundation operated under cooperative agreement by Associated Universities, Inc. This paper makes use of the following ALMA data: ADS/JAO.ALMA\#2019.2.00128.S. ALMA is a partnership of ESO (representing its member states), NSF (USA) and NINS (Japan), together with NRC (Canada), MOST and ASIAA (Taiwan), and KASI (Republic of Korea), in cooperation with the Republic of Chile. The Joint ALMA Observatory is operated by ESO, AUI/NRAO and NAOJ.


\end{acknowledgments}

%






\newpage
\appendix
\section{Double-peak line profiles}\label{app:lines}
\begin{figure*}[h]
\centering
\includegraphics[width=\textwidth]{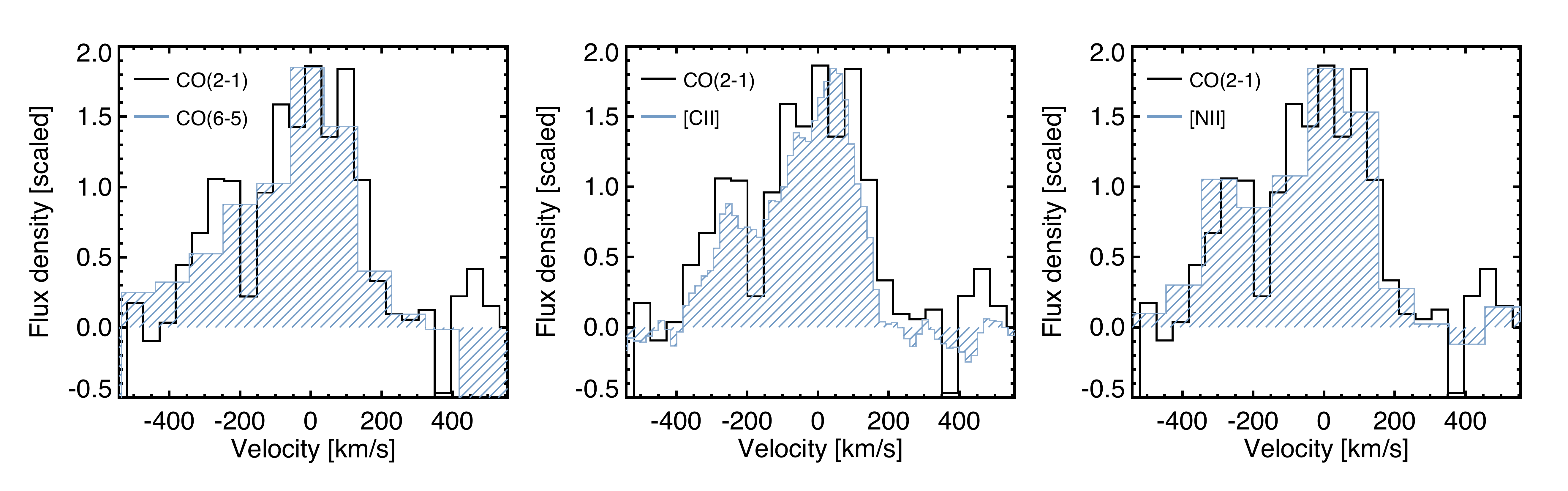}
\caption{{\it Line profiles.} Comparison between the $\rm CO(2\to1)$ line reported in this work and the CO(6-5) from \citealt{Zavala2018a}, the [CII] from the ALMA archive (project code: 2019.2.00128.S), and the [NII] line from \citet{Tadaki2022a}. The lines have been scaled to qualitatively match the line peak of the $\rm CO(2\to1)$ line in order to compare the line profile and line-width. A double-peak line shape is clearly visible in both the atomic and molecular lines.} 
\end{figure*}


\bibliography{sample631}{}
\bibliographystyle{aasjournal}



\end{document}